\documentclass[12pt]{iopart}

\usepackage{iopams}
\usepackage{graphicx}

\usepackage{bm}
\usepackage{amssymb,latexsym}
\usepackage{color}

\begin{document}

\title[Checking the validity of Archimedes' law]{Using surface integrals for checking the Archimedes' law of buoyancy}

\author{F M S Lima}

\address{Institute of Physics, University of Brasilia, P.O. Box 04455, 70919-970, Brasilia-DF, Brazil}

\ead{fabio@fis.unb.br}


\begin{abstract}
A mathematical derivation of the force exerted by an \emph{inhomogeneous} (i.e., compressible) fluid on the surface of an \emph{arbitrarily-shaped} body immersed in it is not found in literature, which may be attributed to our trust on Archimedes' law of buoyancy. However, this law, also known as Archimedes' principle (AP), does not yield the force observed when the body is in contact to the container walls, as is more evident in the case of a block immersed in a liquid and in contact to the bottom, in which a \emph{downward} force that \emph{increases with depth} is observed.  In this work, by taking into account the surface integral of the pressure force exerted by a fluid over the surface of a body, the general validity of AP is checked. For a body fully surrounded by a fluid, homogeneous or not, a gradient version of the divergence theorem applies, yielding a volume integral that simplifies to an upward force which agrees to the force predicted by AP, as long as the fluid density is a \emph{continuous function of depth}. For the bottom case, this approach yields a downward force that increases with depth, which contrasts to AP but is in agreement to experiments. It also yields a formula for this force which shows that it increases with the area of contact.
\end{abstract}



\vspace{0.5cm}

\submitto{\; \EJP}

\vspace{0.5cm}

\noindent
\underline{Accepted for publication}: \, 10/20/2011


\maketitle


\section{Introduction}

The quantitative study of hydrostatic phenomena did begin in antiquity with Archimedes' treatise \emph{On Floating Bodies - Book I}, where some propositions for the problem of the force exerted by a liquid on a body fully or partially submerged in it are proved~\cite{bk:Arch,Graf}.\footnote{Certainly in connection to the need, at that time of flourishing shipping by sea routes, of predicting how much additional weight a ship could support without sinking.}  In modern texts, his propositions are reduced to a single statement known as Archimedes' law of buoyancy, or simply Archimedes' principle (AP), which states that ``\emph{any object immersed in a fluid will experience an upward force equal to the weight of the fluid displaced by the body}''~\cite{teorem,Halliday}. The continuity of this work had to wait about eighteen centuries, until the arising of the scientific method, which was the guide for the experimental investigations in hydrostatics by Stevinus, Galileo, Torricelli, and Pascal, among others.\footnote{Gravesande also deserves citation, due to his accurate experiment for comparing the force exerted by a liquid on a body immersed in it to the weight of the displaced liquid. This experiment uses a bucket and a metallic cylinder that fits snugly inside the bucket. By suspending the bucket and the cylinder from a balance and bringing it into equilibrium, one immerses the cylinder in a water container. The balance equilibrium is then restored by filling the bucket with water.}  The very long time interval from Archimedes to these experimentalists is a clear indicative of the advance of Archimedes thoughts. As pointed out by Netz (based upon a palimpsest discovered recently), Archimedes developed rigorous mathematical proofs for most his ideas~\cite{Netz}. However, the derivation of the exact force exerted by an \emph{inhomogeneous} fluid on an \emph{arbitrarily-shaped} body immersed in it, as will be shown here, demands the knowledge of the divergence theorem, a mathematical tool that was out of reach for the ancients. Therefore, the validity of the Archimedes propositions for this more general case was not formally proved on his original work.

By defining the buoyant force (BF) as the net force exerted by a fluid on the portion of the surface of a body (fully or partially submerged) that touches the fluid, the validity of AP in predicting this force can be checked. In fact, the simple case of symmetric solid bodies (e.g., a right-circular cylinder, as found in Ref.~\cite{Giancoli}, or a rectangular block, as found in Ref.~\cite{Serway}) immersed in a liquid is used in most textbooks for proofing the validity of AP. There, symmetry arguments are taken into account to show that the horizontal forces exerted by the liquid cancel and then the net force reduces to the difference of pressure forces exerted on the top and bottom surfaces~\cite{plausible}.  The BF is then shown to point upwards, with a magnitude that agrees to AP, which also explains the origin of the BF in terms of an increase of pressure with depth~\cite{originBF}.  Note, however, that this proof works only for \emph{symmetric} bodies with horizontal, flat surfaces on the top and the bottom, immersed in an incompressible (i.e., homogeneous) fluid. Although an extension of this result for \emph{arbitrarily-shaped} bodies immersed in a \emph{liquid} can be  found in some textbooks~\cite{Courant,Stewart,MecFlu}, a \emph{formal} generalization for bodies immersed in a \emph{compressible} (i.e., inhomogeneous) fluid is not found in literature. This certainly induces the readers to believe that it should be very complex mathematically, which is not true, as will be shown here.

The existence of exceptions to Archimedes' law has been observed in some simple experiments in which the force predicted by AP is qualitatively \emph{incorrect} for a body immersed in a fluid and in contact to the container walls. For instance, when a symmetric solid (e.g., a cylinder) is fully submerged in a liquid with a face touching the bottom of a container, a \emph{downward} BF is observed, as long as no liquid seeps under the block~\cite{Graf,Bottom,Valiyov,Bierman,Getulio}. Indeed, the experimental evidence that this force \emph{increases with depth} (see, e.g., Refs.~\cite{Graf,Bierman,Mungan,Getulio}) clearly contrasts to the \emph{constant} force predicted by AP. These disagreements led some authors to reconsider the completeness or correctness of the AP statement, as well as the definition of BF itself~\cite{Graf,Valiyov,Bierman,Mungan,australianos}, which seems to make the things more confusing yet.

On aiming at checking mathematically the validity of AP for arbitrarily-shaped bodies immersed in inhomogeneous fluids and intending to elucidate why the bottom case represents an exception to AP, in this work I make use of surface integrals for deriving the exact force exerted by a fluid on the surface of a body immersed in it. When the body does not touch the container walls, a gradient version of the divergence theorem applies, yielding a volume integral that reduces to the weight of the displaced fluid, in agreement to AP.  For the bottom case, this approach yields a force that points \emph{downward} and \emph{increases with depth}, in clear disagreement to AP, but in agreement to experiments. This method also reveals that this force \emph{depends on the area of contact $A_b$ between the body and the bottom}, being equal to the difference between the product $\,p_b \, A_b\,$ ($p_b$ being the pressure at the bottom) and the weight of the displace fluid, a result that is not found in literature and could be explored in undergraduate classes.
\newline

\section{Buoyant force on a body with arbitrary shape}
In modern texts, the Archimedes original propositions are reduced to a single, short statement known as the \emph{Archimedes' principle}~\cite{Halliday,Serway}.

\begin{quotation}
\noindent
{\color{blue} ``When a body is fully or partially submerged in a fluid, a buoyant force $\mathbf{B}$ from the surrounding fluid acts on the body. This force is directed upward and has a magnitude equal to the weight $m_{\!f} \, g$ of the fluid displaced by the body.''}
\end{quotation}

\begin{figure}[pb]
\centering
\scalebox{1.1}{\includegraphics{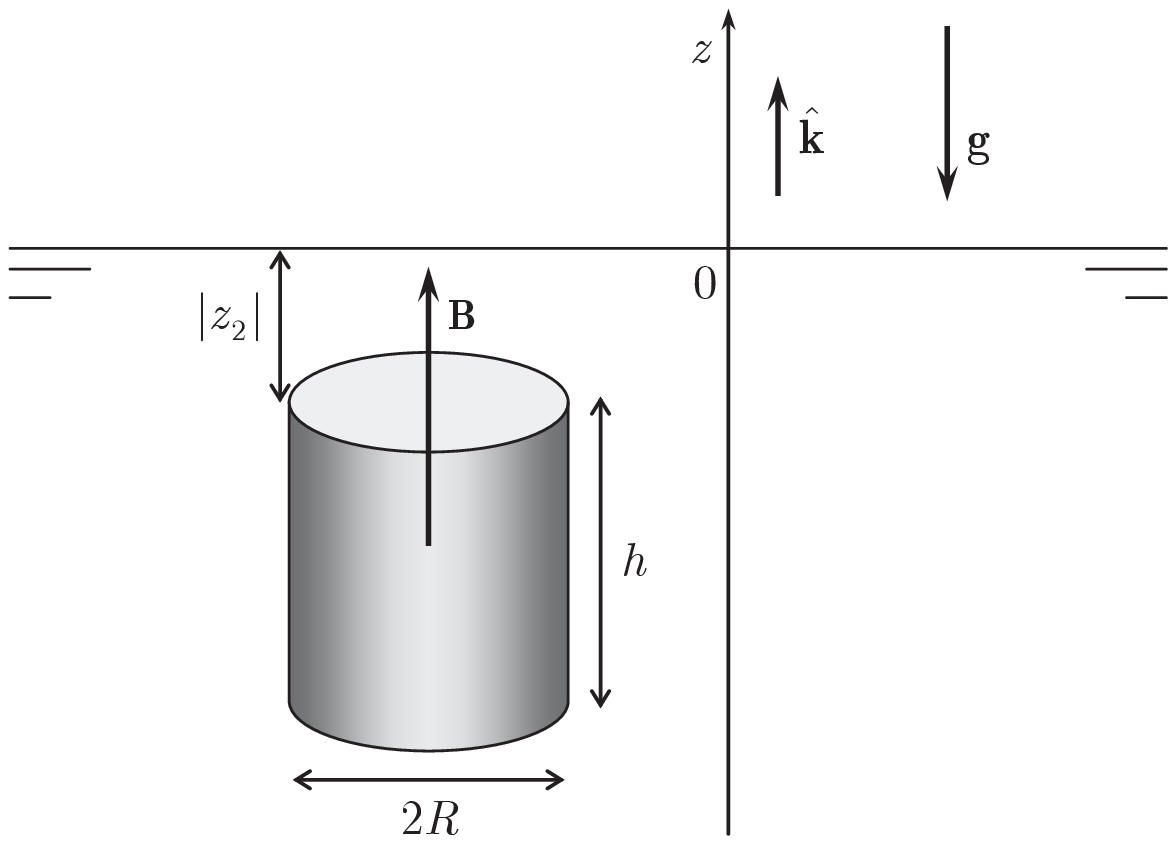}}
\caption{\label{fig:initial}  A cylinder of height $h$ and radius $R$, fully immersed in a liquid. Note that the buoyant force $\mathbf{B}$ is vertical and directed upward. The top surface $S_2$ is kept at a depth $z_2$ below the liquid-air interface (a plane at $z=0$).}
\end{figure}

Here, $m_{\!f}$ is the mass of the fluid that is displaced by the body and $g$ is the local acceleration of gravity.  The BF that follows from AP is simply
\begin{equation}
\mathbf{B} = - \, m_{\!f} \, \mathbf{g} \, ,
\label{eq:BForce}
\end{equation}
where $\mathbf{g} = -g \, \mathbf{\hat{k}}$, $\mathbf{\hat{k}}$ being the unity vector pointing in the $z$-axis direction, as indicated in Fig.~\ref{fig:initial}. For an incompressible, homogeneous fluid --- i.e., a fluid with nearly uniform density~\cite{Liquids} ---, one has $m_{\!f} = \rho \, V_{\!f}$, and then
\begin{equation}
\mathbf{B} = \rho \: V_{\!f} \: g \: \mathbf{\hat{k}}  \, ,
\label{eq:rho}
\end{equation}
where $\rho$ is the density of the fluid and $V_{\!f}$ is the volume of fluid corresponding to $m_{\!f}$.\footnote{Of course, when the body is fully submerged, $V_{\!f}$ equals the volume $V$ of the body.}  In most textbooks, some heuristic arguments similar to the Stevinus ``principle of rigidification'' (see, e.g., Ref.~\cite{rigido}) are taken into account for extending the validity of this formula to \emph{arbitrarily-shaped bodies} immersed in a liquid (see, e.g., Refs.~\cite{Halliday,Giancoli,Serway}), without a mathematical justification. The validity of this generalization is checked below.

Our method starts from the basic relation between the pressure gradient in a fluid in equilibrium in a gravitational field $\mathbf{g}$ and its density $\rho=\rho(\mathbf{r})$, $\mathbf{r}$ being the position vector. From the force-balance for an element of volume of the fluid (homogeneous or not) in equilibrium~\cite{MecFlu}, one finds the well-known \emph{hydrostatic equation}~\cite{MecFlu,bk:atm}:
\begin{equation}
\bm{\nabla} p = \rho(\mathbf{r}) \: \mathbf{g}  \, .
\label{eq:hde}
\end{equation}
For an \emph{uniform}, vertical (downward) gravitational field, as will be assumed hereafter~\cite{campoG}, this simplifies to
\begin{equation}
\bm{\nabla} p = \frac{\partial p}{\partial z} \: \mathbf{\hat{k}} = -\rho(z) \: g \: \mathbf{\hat{k}}  \, .
\label{eq:hde2}
\end{equation}

For a homogeneous (i.e., incompressible) fluid, $\,\rho\,$ is a constant~\cite{Liquids}, and then~\eref{eq:hde2} can be readily integrated, yielding
\begin{equation}
p(z) = p_0 - \rho \, g\, z  \, ,
\label{eq:Stevinus}
\end{equation}
where $p_0$ is the pressure at $z=0$, an arbitrary reference level. This linear decrease of pressure with height is known as \emph{Stevinus law}~\cite{Halliday,Giancoli,Serway}.

Let us now derive a general formula for BF evaluations, i.e. one that works for arbitrarily-shaped bodies, fully or partially submerged in a fluid (or set of distinct fluids) homogeneous or not. For avoiding confusions, let us \emph{define} the buoyant force as \emph{the net force that a fluid exerts on the part $S$ of the surface $\Sigma$ of a body that is effectively in contact to the fluid}. Of course, if the body is fully submerged, the surface $S$ coincides with $\Sigma$. Consider, then, a body immersed in a fluid in equilibrium. Since there is no shearing stress in a fluid at rest, the differential element of force $d \mathbf{F}$ it exerts on a differential element of surface $dS$, at a point $P$ of $S$ in which the fluid touches (and pushes) the body, is normal to $S$ by $P$. Therefore
\begin{equation}
d \mathbf{F} = -p(\mathbf{r}) \: dS \; \mathbf{\hat{n}} \, ,
\label{eq:dF}
\end{equation}
where $\mathbf{\hat{n}} = \mathbf{\hat{n}}(\mathbf{r})$ is the \emph{outward} normal unit vector at point $P$. Note that the pressure force is directed along the \emph{inward} normal to $S$ by $P$.  On assuming that the surface $S$ is piecewise smooth and the vector field $\,p(\mathbf{r}) \: \mathbf{\hat{n}}\,$ is integrable over $S$, one finds a general formula for BF evaluations, namely
\begin{equation}
\mathbf{F} =  -\int\!\!\!\int_S{\: p \; \mathbf{\hat{n}} \; dS} \, .
\label{eq:F}
\end{equation}
This integral can be easily evaluated for a body with a \emph{symmetric} surface fully submerged in a fluid with \emph{uniform density}. For an arbitrarily-shaped body, however, it does not appear to be tractable analytically due to the dependence of the direction of $\mathbf{\hat{n}}$ on the position $\mathbf{r}$ over $S$, which in turn depends on the (arbitrary) shape of $S$. However, this task can be easily worked out for a body fully submerged in a homogeneous fluid ($\rho = $ const.) by applying the divergence theorem to the vector field $\mathbf{E} = -p(z) \, \mathbf{\hat{k}}$, which yields a volume integral that evaluates to $\rho \, g \, V \, \mathbf{\hat{k}}$, in agreement to AP, as discussed in some advanced texts~\cite{Courant,Altintas}.  For the more general case of a body fully or partially submerged in an \emph{inhomogeneous} fluid (or a set of fluids), I shall follow a slightly different way here, based upon the following version of the divergence theorem~\cite{Courant,Kaplan,piecewise}.

\begin{quotation}
\noindent
{\color{blue} \textbf{Gradient theorem.} $\:$ Let $R$ be a bounded region in space whose boundary $S$ is a closed, piecewise smooth surface which is positively oriented by a unit normal vector $\mathbf{\hat{n}}$ directed outward from $R$. If  $f = f(\mathbf{r})$ is a scalar function with continuous partial derivatives in all points of an open region that contains $R$ (including $S$), then
\begin{equation*}
\mathop{\int\!\!\!\!\!\int}_S \mkern-24mu \bigcirc {\: f(\mathbf{r}) \: \mathbf{\hat{n}} \: dS} = \int\!\!\!\int\!\!\!\int_R{\bm{\nabla} \! f \: dV} .
\end{equation*} }
\end{quotation}

In Appendix A, it is shown how the divergence theorem can be used for proofing the gradient theorem.  The advantage of using this less-known calculus theorem is that it allows us to transform the surface integral in Eq.~\eref{eq:F} into a volume integral of $\,\bm{\nabla} p\,$, a vector that can be easily written in terms of the fluid density via the hydrostatic equation, Eq.~\eref{eq:hde2}. As we are only interested in pressure forces, let us substitute  $f(\mathbf{r}) = -p(\mathbf{r})$ in both integrals of the gradient theorem. This yields
\begin{equation}
- \mathop{\int\!\!\!\!\!\int}_S \mkern-24mu \bigcirc {\: p(\mathbf{r}) \: \mathbf{\hat{n}} \: dS} = - \int\!\!\!\int\!\!\!\int_{V_f}{\bm{\nabla} p \: dV} \, ,
\label{eq:tripla}
\end{equation}
where the surface integral at the left-hand side is, according to our definition, the BF itself whenever the surface $S$ is closed, i.e. when the body is \emph{fully submerged} in a fluid. Let us analyze this more closely.

\subsection{A body \emph{fully} submerged in a fluid}

For a body of arbitrary shape fully submerged in a fluid (or a set of fluids), by substituting the pressure gradient in Eq.~\eref{eq:hde2} on Eq.~\eref{eq:tripla}, one finds~\cite{obs1}
\begin{equation}
\mathbf{F} = -\int\!\!\!\int\!\!\!\int_{V_f}{\bm{\nabla} p \: dV} = \left[ \int\!\!\!\int\!\!\!\int_{V_f}{\rho(z) \: dV} \right] g \, \mathbf{\hat{k}} \, .
\label{eq:bfVint}
\end{equation}
For the general case of an inhomogeneous, compressible fluid whose density changes with depth, as occurs with gases and high columns of liquids~\cite{rho_depth}, the pressure gradient in Eq.~\eref{eq:bfVint} will be integrable over $V$ as long as $\rho(z)$ is a continuous function of depth (in conformity to the hypothesis of the gradient theorem). Within this condition, one has
\begin{equation}
\mathbf{F} = \left[ \int\!\!\!\int\!\!\!\int_{V_f}{\rho(z) \: dV} \right] g \: \mathbf{\hat{k}} = \left[ \int\!\!\!\int\!\!\!\int_V{\rho(z) \: dV} \right] g \: \mathbf{\hat{k}} \, .
\end{equation}
Since $\int\!\!\!\int\!\!\!\int_V{\rho(z) \: dV}$ is the mass $\,m_{\!f}\,$ of fluid that would occupy the volume $V$ of the body (fully submerged), then
\begin{equation}
\mathbf{F} = m_{\!f} \, g \; \mathbf{\hat{k}} \, ,
\end{equation}
which is an upward force whose magnitude equals the \emph{weight of the fluid displaced by the body}, in agreement to AP as stated in Eq.~\eref{eq:BForce}. This shows that AP remains valid even for an inhomogeneous fluid, as long as the density is a continuous function of depth, a condition fulfilled in most practical situations.

\subsection{A body \emph{partially} submerged in a fluid}

\begin{figure}[p]
\centering
\scalebox{1.1}{\includegraphics{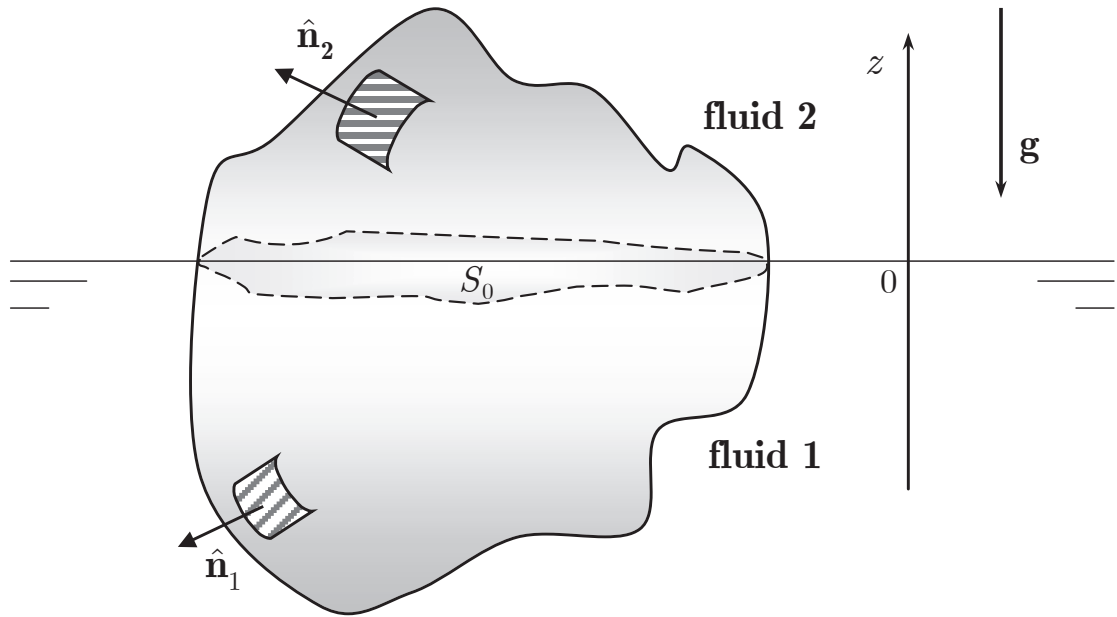}}
\caption{\label{fig:n1n2}  An arbitrarily-shaped body floating in a liquid (fluid 1), with the emerged part in contact to a less dense, compressible fluid (fluid 2). Note that $\mathbf{\hat{n}_1}$ ($\mathbf{\hat{n}_2}$) is the outward unit vector on the surface $S_1$ ($S_2$), as defined in the text. $S_0$ is the horizontal cross-section at the level of the planar interface between fluids 1 and 2 (at $z=0$).}
\end{figure}

The case of an arbitrarily-shaped body floating in a fluid with a density $\rho_1(z)$, with its emerged part exposed to either vacuum (i.e., a fictitious fluid with null density) or a less dense fluid is an interesting example of floating in which the exact BF can be compared to the force predicted by AP.\footnote{A null pressure is assumed on the portions of $S$ that are not interacting with any fluid.} This is important for the study of many floating phenomena, from ships in seawater to the isostatic equilibrium of tectonic plates (known in geology as isostasy)~\cite{bkgeo1}. Without loss of generality, let us restrict our analysis to two fluids, one (denser) with a density $\rho_1(z)$, we call fluid 1, and another (less dense) with a density $\rho_2(z) \le \min{\left[\rho_1(z)\right]}=\rho_1(0^{-})$, we call fluid 2. For simplicity, I choose the origin $z=0$ at the planar surface of separation between the fluids, as indicated in Fig.~\ref{fig:n1n2}, where the fluid density can present a discontinuity $\,\rho_1(0^{-}) -\rho_2(0^{+})$.  The forces that these fluids exert on the body surface can be evaluated by applying the gradient theorem to each fluid separately, as follows. First, divide the body surface $S$ into two parts: the open surface $S_1$ below the interface at $z=0$ and the open surface $S_2$, above $z=0$.  The integral over the (closed) surface $S$ in Eq.~\eref{eq:F} can then be written as
\begin{equation}
\mathop{\int\!\!\!\!\!\int}_S \mkern-24mu \bigcirc {\: p(z) \: \mathbf{\hat{n}} \: dS} = \int\!\!\!\int_{S_1}{\, p(z) \: \mathbf{\hat{n}_1} \: dS} +\int\!\!\!\int_{S_2}{\, p(z) \: \mathbf{\hat{n}_2} \: dS} \, ,
\label{eq:duas}
\end{equation}
where $\mathbf{\hat{n}_1}$ ($\mathbf{\hat{n}_2}$) is the outward unit normal vector at a point of $S_1$ ($S_2$), as indicated in Fig.~\ref{fig:n1n2}. Let us call $S_0$ the planar surface, also indicated in Fig.~\ref{fig:n1n2}, corresponding to the horizontal cross-section of the body at $z=0$. By noting that $\mathbf{\hat{n}_1} = \mathbf{\hat{k}}$ and $\mathbf{\hat{n}_2} = -\mathbf{\hat{k}}$ in all points of $S_0$, then, being $p(z)$ a continuous function, one has
\begin{equation*}
 \int\!\!\!\int_{S_0}{\, p(z) \: \mathbf{\hat{n}_1} \: dS} + \int\!\!\!\int_{S_0}{\, p(z) \: \mathbf{\hat{n}_2} \: dS} = \bm{0} \, .
\end{equation*}
This allows us to use $S_0$ to generate two closed surfaces, $\widetilde{S_1}$ and $\widetilde{S_2}$, formed by the unions $S_1 \cup S_0$ and $S_2 \cup S_0$, respectively. From Eq.~\eref{eq:duas}, one has
\begin{eqnarray}
\mathop{\int\!\!\!\!\!\int}_S \mkern-24mu \bigcirc {\: p(z) \: \mathbf{\hat{n}} \: dS} &=&  \int\!\!\!\int_{S_1}{\, p(z) \: \mathbf{\hat{n}_1} \: dS} + \int\!\!\!\int_{S_0}{\, p(z) \: \mathbf{\hat{n}_1} \: dS} \nonumber \\
&+&\int\!\!\!\int_{S_2}{\, p(z) \: \mathbf{\hat{n}_2} \: dS} +  \int\!\!\!\int_{S_0}{\, p(z) \: \mathbf{\hat{n}_2} \: dS} \nonumber \\
&=& \mathop{\int\!\!\!\!\!\int}_{\widetilde{S_1}} \mkern-24mu \bigcirc {\: p(z) \: \mathbf{\hat{n}_1} \: dS} + \mathop{\int\!\!\!\!\!\int}_{\widetilde{S_2}} \mkern-24mu \bigcirc {\: p(z) \: \mathbf{\hat{n}_2} \: dS} \, .
\label{eq:int4}
\end{eqnarray}
As both $\widetilde{S_1}$ and $\widetilde{S_2}$ are closed surfaces, one can apply the gradient theorem to each of them, separately. This gives
\begin{eqnarray}
\mathbf{F} &=& - \left(\mathop{\int\!\!\!\!\!\int}_{\widetilde{S_1}} \mkern-24mu \bigcirc {\: p(z) \: \mathbf{\hat{n}_1} \: dS} + \mathop{\int\!\!\!\!\!\int}_{\widetilde{S_2}} \mkern-24mu \bigcirc {\: p(z) \: \mathbf{\hat{n}_2} \: dS} \right) \nonumber \\
&=&  - \left(  \int\!\!\! \int\!\!\!\int_{V_1}{\bm{\nabla} p \: dV} +  \int\!\!\! \int\!\!\!\int_{V_2}{\bm{\nabla} p \: dV} \right) \nonumber \\
&=&  - \left(  \int\!\!\!\int\!\!\!\int_{V_1} {\frac{{\partial p}}{{\partial z}}\:dV} + \int\!\!\!\int\!\!\!\int_{V_2} {\frac{{\partial p}}{{\partial z}}\:dV} \right) \mathbf{\hat{k}} \, ,
\label{eq:soma0}
\end{eqnarray}
where $V_1$ and $V_2$ are the volumes of the portions of the body below and above the interface at $z=0$, respectively. From the hydrostatic equation, one has ${\,{\partial p}/{\partial z}} = -g\,\rho(z)$, which reduces the above integrals to
\begin{eqnarray}
\int\!\!\!\int\!\!\!\int_{V_1} {\left[ - \rho_1(z) \,g \right]\,dV} + \int\!\!\!\int\!\!\!\int_{V_2} {\left[ - \rho_2(z) \,g \right]\,dV} \nonumber \\
= -g \left[\int\!\!\!\int\!\!\!\int_{V_1} {\rho_1(z)\:dV} +\int\!\!\!\int\!\!\!\int_{V_2} {\rho_2(z)\:dV} \right] .
\label{eq:antessoma}
\end{eqnarray}
The latter volume integrals are equivalent to the masses $m_1$ and $m_2$ of the fluids 1 and 2 displaced by the body, respectively, which reduces the BF to
\begin{equation}
\mathbf{F} = g \left[ \int\!\!\!\int\!\!\!\int_{V_1} {\rho_1(z)\:dV} +\int\!\!\!\int\!\!\!\int_{V_2} {\rho_2(z)\:dV} \right] \mathbf{\hat{k}} = \left( m_1 + m_2 \right) \, g \, \mathbf{\hat{k}} \, .
\label{eq:soma}
\end{equation}
The BF is then upward and its magnitude is equal to the sum of the weights of the fluids displaced by the body, in agreement to AP in the form given in Eq.~\eref{eq:BForce}. Note that the potential energy minimization technique described in Refs.~\cite{Leroy,Vermi,Reed} cannot provide this confirmation of AP because it works only for rigorously \emph{homogeneous} (i.e., incompressible) fluids. Interestingly, our proof shows that the exact BF can also be found by assuming that the body is fully submerged in a \emph{single} fluid with a variable density $\rho(z)$ that is not continuous, but a \emph{piecewise} continuous function with a (finite) leap discontinuity at $z=0$.\footnote{Interestingly, this suggests that a more general version of the divergence theorem could be found, in which the requirement of continuity of the partial derivatives of $f(\mathbf{r})$ (respectively, of $\bf{\nabla} \cdot \mathbf{E}$) could be weakened to only a \emph{piecewise continuity}. I have not found such generalization in literature.}

Although the contribution of fluid 2 to the BF is usually smaller than that of fluid~1, it cannot in general be neglected, as done in introductory physics textbooks~\cite{Lan}.  In our approach, this corresponds to assume a constant pressure on all points of the surface $S_2$ of the emerged portion, which is incorrect. This leads to a null gradient of pressure on the emerged part of the body, which erroneously reduces the BF to only
\begin{equation}
\mathbf{F} = \left[ \int\!\!\!\int\!\!\!\int\limits_{V_1}{ \rho_1(z) \,dV} \right] g \: \mathbf{\hat{k}} \, .
\label{eq:naive}
\end{equation}
Being fluid 1 a liquid, as usual, then $\rho_1(z)$ is nearly a constant (let us call it $\rho_1$), which simplifies this upward force to $\rho_1 \, V_1 \,g$. This is the result presented for the water-air pair in most textbooks~\cite{Halliday,Giancoli,Serway}. It is also the result that can be deduced from Archimedes original propositions~\cite{Rorres}, as well as the result found by minimizing the potential energy~\cite{Leroy,Vermi,Reed}. It is clear that this naive approximate result always \emph{underestimates} the actual BF established in Eq.~\eref{eq:soma}, being a reasonable approximation only when $m_2 \ll m_1$~\cite{densities}. The inclusion of the term corresponding to $m_2$ is then essential for accurate evaluations of the BF, as shown by Lan for a block floating in a liquid~\cite{Lan}.  For an arbitrarily-shaped body immersed in a liquid-gas fluid system, our Eq.~\eref{eq:antessoma} yields the following expression for the exact BF:
\begin{equation}
\mathbf{F} = \left( \rho_1 \,g\, V_1 - \int\!\!\!\int\!\!\!\int_{V_2} {\frac{{\partial p}}{{\partial z}} \: dV} \right) \mathbf{\hat{k}} \, .
\label{eq:acima}
\end{equation}
Therefore, the function $p(z)$ on the emerged portion of the body determines the contribution of fluid 2 to the exact BF. If fluid 2 is approximated as an incompressible fluid, i.e. if one assumes $\rho_2(z) \approx \rho_2(0) \equiv \rho_2$, by applying the Stevinus law one finds
\begin{equation}
\mathbf{F} \approx \left\{ \rho_1 \,g\, V_1 - \int\!\!\!\int\!\!\!\int_{V_2} {\left[ -\rho_2(0)\,g \right]\:dV} \right\} \, \mathbf{\hat{k}} = \left(\,\rho_1 \,g\,V_1 + \rho_2\,g\, V_2 \right) \mathbf{\hat{k}} \, .
\label{eq:rho20}
\end{equation}
This is just the result found by Lan by applying AP in the form stated in our Eq.~\eref{eq:rho} to fluids 1 and 2, separately, and then summing up the results~\cite{Lan}. Of course, this approximation is better than Eq.~\eref{eq:naive}, but, contrarily to Lan's opinion, the \emph{correct} result arises only when one takes into account the decrease of $\rho_2$ with $z$.  This demands the knowledge of a barometric law, i.e. a formula for $p(z)$ with $z > 0$. Fortunately, most known barometric laws are derived just from the hydrostatic equation~\cite{bk:atm}, which allows us to substitute the pressure derivative in Eq.~\eref{eq:acima} by $\,-\rho_2(z)\,g$, finding that
\begin{eqnarray}
\mathbf{F} &=& \left[ \rho_1 \,g\, V_1 - \int\!\!\!\int\!\!\!\int_{V_2}{\left[-\rho_2(z)\,g\right] \: dV} \right] \mathbf{\hat{k}} \nonumber \\
&=& \left[ \rho_1 \, V_1 +\int\!\!\!\int\!\!\!\int_{V_2}{\rho_2(z) \: dV} \right] g \; \mathbf{\hat{k}} \, ,
\label{eq:best}
\end{eqnarray}
which promptly reduces to the exact result found in Eq.~\eref{eq:soma}.\footnote{When fluid 2 is vacuum --- i.e., the body is \emph{partially submerged} (literally) in fluid 1 only --- the exact result in Eq.~\eref{eq:best}, above, is indeed capable of furnishing the correct force (namely, an upward force with a magnitude $\,\rho_1 \, V_1 \, g$), which is found by taking the limit as $\rho_2(z)$ tends to zero uniformly.}

\section{Exceptions to Archimedes' principle}

\begin{figure}[pt]
\centering
\scalebox{1.1}{\includegraphics{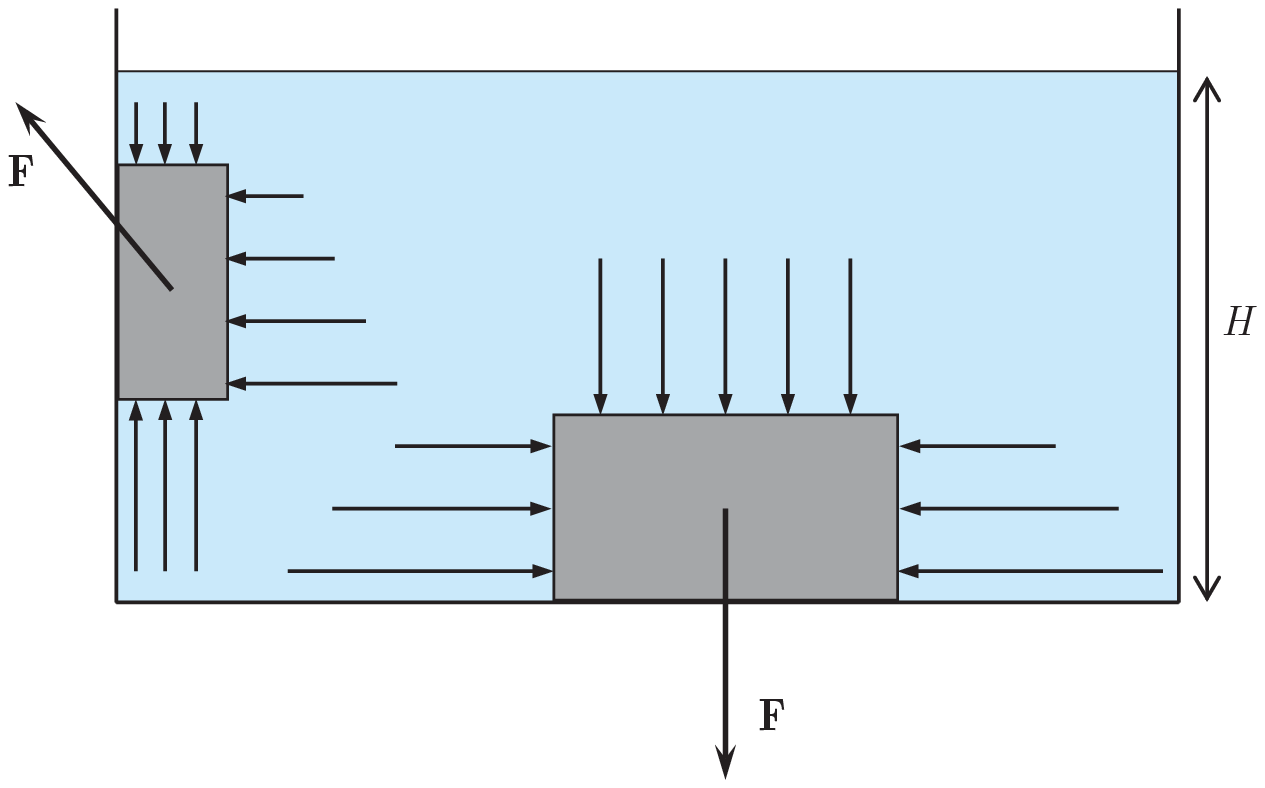}}
\caption{\label{fig:bottom}  The hydrostatic forces acting on rectangular blocks in contact to the walls of a container. The arrows indicate the pressure forces exerted by the liquid on the surface of each block. The net force exerted by the liquid in each block, i.e. the `buoyant' force (as defined in the text), is represented by the vector $\mathbf{F}$. The larger block represents the \emph{bottom case}.}
\end{figure}

When a body is immersed in a liquid and put in contact to the container walls, as illustrated in Fig.~\ref{fig:bottom}, a buoyant force is observed which does not agree to that predicted by AP.  This has been observed in some simple experiments in which a symmetric solid (typically a cylinder or a rectangular block) is fully submerged in a liquid with a face touching the bottom of a container.  Let us call this the \emph{bottom case}~\cite{Bottom}. In this case, a \emph{downward} BF (according to our definition) is observed if no liquid seeps under the block~\cite{Bottom,Valiyov,Bierman,Getulio}. There is indeed experimental evidence that the magnitude of this force \emph{increases linearly with depth} (see, e.g., Refs.~\cite{Graf,Bierman,Getulio}), which contrasts to the \emph{constant} force ($= \rho \, V g \: \mathbf{\hat{k}}$) predicted by AP. These experimental results have led some authors to reconsider the completeness or correctness of the AP statement, as well as the definition of buoyant force itself~\cite{Graf,Valiyov,Bierman,Mungan,australianos}.

In spite of the experimental difficulties in studying these exceptions to AP, I shall apply the surface integral approach to determine the exact BF that should be observed in an ideal experiment in which there is no fluid under the block. This will help us to understand why the bottom case represents an exception to AP.  For a symmetric solid body (either a cylinder or a rectangular block) with its flat bottom resting on the bottom of a container, as illustrated for the block at the right of Fig.~\ref{fig:bottom}, if no liquid seeps under the block then the horizontal forces cancel and the \emph{net force exerted by the liquid} reduces to the downward pressure force exerted by the liquid on the top surface~\cite{Bottom,Valiyov,Bierman}. From Stevinus law, $p_{top} = p_0 + \rho\,g\,\left| z_{top} \right|$, $z_{top}$ being the depth of the top, therefore
\begin{equation}
\mathbf{F} = -\int\!\!\!\int_{S_{top}}{p(z) \, \mathbf{\hat{n}} \: dS} = -p_{top} \: A \: \mathbf{\hat{k}} = -\left( p_0 + \rho \, g \, \left|z_{top}\right| \right) A \: \mathbf{\hat{k}} \, ,
\label{eq:dwbf1}
\end{equation}
where $A$ is the area of the top surface. This \emph{downward} force then \emph{increases linearly with depth}, which clearly contrasts to the force predicted by AP, but agrees to experimental results~\cite{Graf,Bierman,Getulio}. In fact, the increase of this force with depth has been subject of deeper discussions in recent works~\cite{Graf,Bierman,Mungan,Kibble}, in which it is suggested that the meaning of the word `immersed' should be `fully surrounded by a liquid' instead of `in contact to a liquid', which would make the `bottom' case, as well as all other `contact cases', out of scope of the Archimedes original propositions, as well as AP modern statement~\cite{Graf,Bierman,Kibble}. Note, however, that this redefinition is deficient because it excludes some common cases of buoyancy such as, for instance, that of a solid (e.g., a piece of cork) floating in a denser liquid (e.g., water). In this simple example, the body is not fully surrounded by a liquid and yet AP works!  More recently, other authors have argued that the definition of BF itself should be changed to ``an upward force with a magnitude equal to the weight of the displaced fluid''~\cite{Mungan}. However, I have noted that this would make AP a \emph{definition} for the BF and then, logically, AP would not admit any exceptions at all. In face of the \emph{downward} BF experiments already mentioned, it is clear that this is not a good choice of definition. Therefore, I would like to propose the abandon of such redefinitions, as they are unnecessary once we admit some exceptions to the AP, which is the natural way to treat the \emph{exceptional} cases not realized by Archimedes in his original work.

For a better comparison to the result for arbitrarily-shaped bodies that will be derived below, let us write the force found in Eq.~\eref{eq:dwbf1} in terms of the pressure $p_b$ at the bottom of the container. As the reader can easily check, this yields
\begin{equation}
\mathbf{F} = - \left(p_b \, A  -\rho \, V g \right) \, \mathbf{\hat{k}} \, .
\label{eq:dwbf2}
\end{equation}
Note that this simple result is for a \emph{`vacuum' contact}, i.e. an ideal contact in which neither liquid nor air is under the block. This is an important point to be taken into account by those interested in to develop a downward BF experiment similar to that proposed by Bierman and Kincanon~\cite{Bierman}, since a part of the bottom of the block with an area $A_{air}$ is intentionally left in contact to air (this comes from their technique to reduce the liquid seepage under the block)~\cite{Getulio}. This changes the BF to
\begin{equation}
\mathbf{F} = -\left( p_b\,A - p_0\,A_{air} -\rho \, V g \right) \, \mathbf{\hat{k}} \, .
\label{eq:dwbfAIR}
\end{equation}

For an \emph{arbitrarily-shaped body} immersed in a liquid (in the bottom case), let us assume that there is a non-null area $A_b$ of direct contact between the body and the bottom of a container. If no liquid seeps under the block, then the pressure exerted by the \emph{liquid} there at the bottom of the body is of course null. The BF is then
\begin{eqnarray*}
\mathbf{F} &=& -\int\!\!\!\int_{S_2 \cup A_b}{p(z) \, \mathbf{\hat{n}} \: dS} = -\left[\int\!\!\!\int_{S_2}{p(z) \, \mathbf{\hat{n}} \: dS} +\int\!\!\!\int_{A_b}{0 \, \left( -\mathbf{\hat{k}} \right) \: dS} \right] \\
&=& -\int\!\!\!\int_{S_2}{p(z) \, \mathbf{\hat{n}} \: dS} \, .
\label{eq:dwbf3}
\end{eqnarray*}
In view to apply the gradient theorem, one needs a surface integral over a \emph{closed} surface. By creating a fictitious closed surface $\Sigma = S_2 \cup A_b$ on which the pressure forces will be exerted as if the body would be fully surrounded by the liquid (i.e., one assumes a constant pressure $p_b$ over the horizontal surface $A_b$), one has
\begin{eqnarray}
\mathbf{F} &=& -\int\!\!\!\int_{S_2}{p(z) \, \mathbf{\hat{n}} \: dS} -\int\!\!\!\int_{A_b}{p_b \, \left(-\mathbf{\hat{k}}\right) \, dS} +\int\!\!\!\int_{A_b}{p_b \, \left(-\mathbf{\hat{k}}\right) \, dS} \nonumber \\
&=& -\left[\mathop{\int\!\!\!\!\!\int}_{\Sigma} \mkern-24mu \bigcirc {\: p(z) \, \mathbf{\hat{n}} \: dS} -\int\!\!\!\int_{A_b}{p_b\,\left( -\mathbf{\hat{k}} \right) dS} \right] = -\mathop{\int\!\!\!\!\!\int}_{\Sigma} \mkern-24mu \bigcirc {\: p(z) \, \mathbf{\hat{n}} \: dS} -p_b\,\mathbf{\hat{k}}\,\int\!\!\!\int_{A_b}{dS} \nonumber \\
&=& \rho\,V\,g\,\mathbf{\hat{k}} -p_b\,A_b\,\mathbf{\hat{k}} = -\left(p_b\,A_b -\rho \, V g \right) \, \mathbf{\hat{k}} \, .
\label{eq:dwbf4}
\end{eqnarray}
This is again a \emph{downward} BF that increases linearly with depth, since $p_b=p_0+\rho\,g\,H$, $H$ being the height of the liquid column above the bottom, as indicated in Fig.~\ref{fig:bottom}. This result for arbitrarily-shaped bodies is not found in literature.  Incidentally, this result suggest that the only exceptions to AP, for fluids in equilibrium, are those cases in which $\bm{\nabla} p$ is not a piecewise continuous function over the whole surface $\Sigma\,$ of the body, otherwise the results of the previous section guarantee that the BF points upward and has a magnitude $\,\rho\,V g$, in agreement to AP. This includes all contact cases, since the boundary of the contact region is composed by points around which the pressure leaps (i.e., changes discontinuously) from a strictly positive value $\,p(z)\,$ (at the liquid side) to an smaller (ideally null) pressure (at the contact surface). Therefore, in these points the pressure is not a differentiable function (because it is not even a continuous function), which impedes us of applying the gradient theorem~\cite{ultimaRef}.

\section{Conclusions}

Here in this paper, I have drawn the attention of the readers to the fact that the BF predicted by AP can be derived mathematically even for bodies of arbitrary shape, fully or partially submerged in a fluid, homogeneous or not, based only upon the validity of the hydrostatic equation and the gradient theorem. For that, I first define the buoyant force as the net force exerted by a fluid on the portion of the surface of the body that is pressed by the fluid. Then, the exact BF becomes a surface integral of the pressure force exerted by the fluid, which can be easily evaluated when the body is fully surrounded by the fluid. In this case, the gradient theorem allows one to convert that surface integral into a volume integral which promptly reduces to an upward force with a magnitude equal to the weight of the displaced fluid (as predicted by AP), as long as the fluid density is a \emph{continuous function of depth}.

Finally, some cases were pointed out in which AP fails and this could help students (even teachers) to avoid erroneous applications of this physical law. The exact force in one of these exceptional cases is determined here by applying our surface integral approach to a body immersed in a liquid and in contact to the bottom of a container. In this case, our result agrees to some recent experiments in which it is shown that the force exerted by the liquid is a \emph{downward} force that \emph{increases linearly with depth}, in clear contrast to the force predicted by AP. The method introduced here is indeed capable of providing a formula for the correct force, valid for bodies with arbitrary shapes, which involves the area of contact. Since Archimedes was one of the greatest geniuses of the ancient world, it would not be surprising that he had enunciated his theorems with remarkable precision and insight, however there are some instances he did not realize. These cases are shown here to be exceptions to the AP, thus it would be insensate to make great efforts to keep AP valid without exceptions at the cost of deficient redefinitions.

Since the method presented here is not so complex mathematically, involving only basic rules of vector calculus, it could be included or mentioned in textbooks, at least in the form of a reference that could be looked up by the more interested readers.
\newline

\section*{Appendix A: Proof of the gradient theorem}

Let us show how the Gauss's divergence theorem (see, e.g., Refs.~\cite{Courant,Stewart,Kreyszig,Kaplan}) can be applied to proof the gradient theorem.

{\color{blue}
\begin{quotation}
\noindent
\textbf{Divergence theorem.}  Suppose that $R$ and $S$ satisfy the conditions mentioned in the gradient theorem. If $\,\mathbf{E}=\mathbf{E}(\mathbf{r})$ is a vector field whose components have continuous partial derivatives in all points of $\,V$ (including $S$), then
\begin{equation*}
\mathop{{\int\!\!\!\!\!\int}\mkern-21mu \bigcirc}\limits_S {{\mathbf{E}} \cdot {\mathbf{\hat{n}}}\:dS}  = \int\!\!\!\int\!\!\!\int\limits_R {\bm{\nabla}  \cdot {\mathbf{E}}\:dV} .
\end{equation*}
\end{quotation}  }

\noindent
\emph{Proof (of gradient theorem).} \; Let $P$ be a point over the closed surface $S$ that bounds $R$. Suppose that $f=f(\mathbf{r})$ has continuous partial derivatives at every point in $R$, including those at $S$. By choosing $\mathbf{E} = f \, \mathbf{c}$, $\mathbf{c} \ne \mathbf{0}$ being an arbitrary constant vector, and substituting it in the integrals of the divergence theorem, above, one finds
\begin{equation}
\mathop{{\int\!\!\!\!\!\int}\mkern-21mu \bigcirc}\limits_S
 {\left( {f\,{\mathbf{c}}} \right) \cdot {\mathbf{\hat{n}}}\,dS}  = \int\!\!\!\int\!\!\!\int\limits_R{\bm{\nabla}  \cdot \left( {f\,{\mathbf{c}}} \right)\,dV} \, .
\end{equation}
Since $\bm{\nabla}  \cdot {\mathbf{c}} = 0$, then $\,\bm{\nabla}  \cdot \left( {f\,{\mathbf{c}}} \right) = f\left( {\bm{\nabla}  \cdot {\mathbf{c}}} \right) + {\mathbf{c}} \cdot \bm{\nabla} f = {\mathbf{c}} \cdot \bm{\nabla} f$. Therefore
\begin{equation*}
\mathop{{\int\!\!\!\!\!\int}\mkern-21mu \bigcirc}\limits_S {{\mathbf{c}} \cdot \left( {f\,{\mathbf{\hat{n}}}} \right)\,dS}  = \int\!\!\!\int\!\!\!\int\limits_R {{\mathbf{c}} \cdot \left( {\bm{\nabla} f} \right)\,dV} \, ,
\end{equation*}
which implies that
\begin{equation*}
{\mathbf{c}} \cdot \left( {\mathop{{\int\!\!\!\!\!\int}\mkern-21mu \bigcirc}\limits_S
 {f\,{\mathbf{\hat{n}}}\,dS}  - \int\!\!\!\int\!\!\!\int\limits_R {\bm{\nabla} f\,dV}} \right) = 0 \, .
\end{equation*}
By hypothesis, $\mathbf{c} \ne \mathbf{0}$. If the vector into parentheses, above, were not null, it should always be perpendicular to $\mathbf{c}$, in order to nullify the scalar product, which is impossible because $\mathbf{c}$ is an arbitrary vector. Therefore, one has to conclude that
\begin{equation*}
\mathop{{\int\!\!\!\!\!\int}\mkern-21mu \bigcirc}\limits_S {f\,{\mathbf{\hat{n}}}\,dS} = \int\!\!\!\int\!\!\!\int\limits_R {\bm{\nabla} f\,dV} . \qquad \qquad \qquad \qquad \qquad \qquad \qquad \Box
\end{equation*}


\quad \newline


\begin{thebibliography}{99}
\bibitem{teorem} Note that Archimedes did not call his discoveries in hydrostatics by ``laws,'' nor did presented them as a consequence of experiments. Instead, he treated them as mathematical theorems, similarly to those proposed by Euclides for geometry.
\bibitem{plausible} Of course, the only physically plausible cause for this upward force is the greater fluid pressure on the bottom of the cylinder in comparison to the pressure on the top.
\bibitem{originBF} The origin of the BF is not found in the Archimedes' original work.
\bibitem{rho_depth} Being rigorous, all fluids present some increase of density with depth, but the increasing rate for liquids is much smaller than that for gases.
\bibitem{densities} In fact, for most applications involving a liquid-air pair the second term in Eq.~\eref{eq:rho20} represents only a small correction to the naive approximation in Eq.~\eref{eq:naive} because air density is generally much smaller than typical liquid densities. For instance, at sea level, air density is only $1.2$~kg/m$^3$ whereas fresh water density is about $1,000$~kg/m$^3$.
\bibitem{campoG} The assumption of an uniform gravitational field is a very good approximation for points near the Earth's surface. However, for problems involving large depths/heights of fluids (e.g., undersea extraction of oil, isostasy in deep crustal layers, and others), the decay of the gravitational field with the distance to the center of Earth has to be taken into account. For these applications, a more accurate model (reconsidering Eqs.~\eref{eq:hde2} and~\eref{eq:Stevinus}) has to be developed.
\bibitem{obs1} For a homogeneous fluid, one has $\rho(\mathbf{r}) = \rho =$ const., which reduces the integral in Eq.~\eref{eq:bfVint} to
$\rho \, g \left( \int\!\!\!\int\!\!\!\int_{V_f}{\: dV} \right) \mathbf{\hat{k}} = \rho \, g \, V_f \, \mathbf{\hat{k}} = \rho \, g \, V \, \mathbf{\hat{k}}$, which, as expected, agrees to AP.
\end{thebibliography}
\end{document}